\documentclass{fmcad}

\usepackage{common}
\usepackage{xspace}
\usepackage{graphics,subcaption,listings} 
\title{Nazrin: An Atomic Neural Proof Automation Tactic in Lean 4}

\begin{document}

\author{
	\IEEEauthorblockN{Leni Aniva \orcid{0000-0002-6033-9140}}
	\IEEEauthorblockA{Stanford University, Stanford, U.S. \\ aniva@stanford.edu}
	\and
	\IEEEauthorblockN{Iori Oikawa \orcid{0000-0001-8385-6901}}
	\IEEEauthorblockA{Northeastern University, Massachusetts, U.S. \\ wang.fengk@northeastern.edu}
	\and
	\IEEEauthorblockN{David Dill \orcid{0000-0002-6189-0866}}
	\IEEEauthorblockA{Stanford University \\ dill@cs.stanford.edu}
	\and
	\IEEEauthorblockN{Clark Barrett \orcid{0000-0002-9522-3084}}
	\IEEEauthorblockA{Stanford University \\ barrettc@stanford.edu}
}
%
%
%

\maketitle

\begin{abstract}
	In Machine-Assisted Theorem Proving, a theorem proving agent searches for a
sequence of expressions and tactics that can prove a statement in a proof
assistant.
In this work, we introduce several novel concepts and capabilities to address
obstacles faced by machine-assisted theorem proving.  We first present a set of
\keyword{atomic tactics}, a small finite set of tactics capable of proving any
provable statement in Lean.  We then introduce a \keyword{transposing
atomization} algorithm which turns arbitrary proof expressions into a series of
atomic tactics.  We next introduce the \keyword{ExprGraph} data structure, which
provides a succinct representation for Lean expressions.  Finally, we present
the \keyword{Nazrin Prover}, short for \textbf{N}eural
\textbf{A}tomi\textbf{z}e\textbf{r} for \textbf{In}habitation Problems, a graph
neural network-based theorem proving agent using atomic tactics and ExprGraph.
Nazrin circumvents many challenges faced by existing proving agents by
exclusively dispatching atomic tactics, and it is robust enough to both train
and evaluate on consumer-grade hardware.  We demonstrate the potential of tools
like Nazrin using theorems from Lean's standard library and from Mathlib.

\end{abstract}

\section{Introduction}
Most modern mathematical proofs are still written in natural language, a representation
susceptible to ambiguity, errors, and logic gaps. A \textbf{proof assistant} is a computer
program that implements a formal proof system for foundational mathematical
reasoning.  Using a proof assistant, we can write formal mathematical proofs
that are precise and are guaranteed to have no gaps or errors. \textbf{Lean 4}
(referred to simply as Lean below, for brevity) is a widely used proof assistant
\cite{lean4} based on the Calculus of Inductive Constructions. Its
Mathlib \cite{mathlib} library is one of the largest corpora of formalized
mathematics ever built. In Lean, the type of a theorem is its statement, and the
value of a theorem is its proof. This Curry-Howard correspondence between
theorems and functions means that the problem of finding a proof is equivalent
to the problem of inhabitation: finding a member of a type. Proofs in Lean are
typically written manually as a series of tactics and expressions that decompose
and discharge proof obligations.

\Comment{DD: Formalization is irrelevant to the paper.  Also, talking about it in the 2nd
paragraph of the intro sets an expectation that it's an important part of the paper.}
\Comment{DD: Not sure we need the term MATP.}
Interactive theorem proving is laborious and can be tedious because the user
must deal with every single detail of lengthy formal proofs.  To mitigate this, modern
proof assistants incorporate a collection of automated proof methods to do much of the
work, especially low-level ``obvious'' proofs.  These methods include decision
procedures for fragments of logic~\cite{leanauto}
\cite{leansmt}, resolution theorem provers, hand-coded heuristic
search \cite{aesop}, or the Hammer-like
Grind tactic in Lean \cite{lean4}. There is a clear opportunity to apply machine
learning to learn heuristics from the extensive library of existing proofs,
instead of hand coding them.

A natural approach is to train an agent on human-written proofs.  However,
this approach has several drawbacks. First of all, Lean has an unbounded space of
tactics that can be used to prove the same thing in many different ways,
and existing proofs use arbitrary collections of these tactics. This leads to a
noisy data set, where it is unclear whether the choice of the next tactic is because
it's the obvious right thing to do, or because that's what the author happened to prefer.
We would expect this to  confuse or slow down training.  Second, the final version of a proof written
by a human typically emphasizes conciseness and ease of checking, and does not provide much insight
about how to \textit{find} a proof. Third, expressions in Lean contain
type-theoretically immaterial information to the mathematical argument which
could be elided without modifying the validity of a proof.

In this paper, we introduce a novel proof search approach which addresses these
drawbacks.  We have five main contributions:
\begin{enumerate}
\item \textbf{Atomic Tactics}: We introduce a small set of simple tactics with a
bounded parameter space, which can be used to incrementally construct a proof expression in
Lean. This provides a small, finite action space for theorem proving agents,
reducing the number of choices that have to be made at each step of proof
search.

	\Comment{DD: Is it appropriate to mention RL here?  If we're even using RL.}
\item \textbf{Transposing Atomization}: We present an algorithm that can translate
	a proof term into a series of atomic tactics. This can be used
	to generate a large amount of training data from existing formalized
	mathematics. The generated data trains the system \textit{how to find proofs},
	not how to present them. The method of atomization could in principle be
	implemented in any proof assistant.
\item \textbf{ExprGraph}: We design a reduced graph-based representation of
	expressions in Lean that exploits the fundamental symmetries of mathematics,
	while preserving a necessary amount of information for search to proceed.
	We explicitly construct variable reference edges in the ExprGraph to mimic
	attention mechanisms present in language models.
\item \textbf{Nazrin Prover}: We introduce \textbf{Nazrin}, short for \textbf{N}eural
\textbf{A}tomi\textbf{z}e\textbf{r} for \textbf{In}habitation Problems, a
	low-resource graph neural network-based theorem proving agent which is
	trained, using atomized data and reinforcement learning, to output
	atomized tactics. Nazrin has similar aims to
	proof automation methods like Sledgehammer \cite{sledgehammer} (in
	Isabelle/HOL) and Aesop	\cite{aesop} in Lean.
\item \textbf{Evaluation}: We evaluate Nazrin on Lean's standard library and on
	its power to generalize on Mathlib. This mimics the use case of developing a
	new theory in Mathlib, which requires the prover agent to generalize on
	Mathlib itself. 
\end{enumerate}

The remainder of the paper is organized as follows.  We start with related work
(\ref{sec:Related}) and background (\ref{sec:Background}).  We then introduce
our atomic tactics and atomization algorithm (\ref{sec:Methodology}).  We next
explain our graph-based representations of Lean expressions and how they are
used as core building blocks for the Nazrin prover (\ref{sec:Graph}).  Finally,
we present our evaluation (\ref{sec:Evaluation}) and conclude
(\ref{sec:Conclusion}).

\section{Related Work}\label{sec:Related}
Tree search, e.g., MCTS~\cite{mcts1987}, is a common paradigm for tasks involving
decisions (e.g., Board games, path planning, theorem proving, etc).  Previous
work on using tree search for theorem proving includes \cite{hypertree},
BFS-Prover \cite{bfs-prover}, and DT-Solver \cite{wang2023dt}.  Hypertree Proof
Search in particular pioneered the use of product rewards for determining the
estimated reward for multiple goals. It uses an And-Or tree structure to
represent the search tree. BFS-Prover \cite{bfs-prover} uses a language model to
estimate the value of a goal. DT-Solver \cite{wang2023dt} likewise uses a LLM
but also estimates the viability of a goal by comparing it against its
ancestors, which uses LLMs to perform tree search. Alternatively, an LLM can
interact with a proof assistant directly via text \cite{deepseek-prover}.
LeanDojo and ReProver \cite{leandojo} involve using a language model as a
reinforcement learning agent whose objective is to discharge all goals with
tactics.

Previous work on using graphs to represent expressions includes
\cite{graph4hol}, in which GNNs are used to perform premise
selection in HOL, which is similar to Lean. Graphs can also be used for premise
selection, e.g., in \cite{graph-premise-selection}. Other methods for exploiting
symmetries in mathematics include de Bruijn indices \cite{deBruijnNames}.

Aesop \cite{aesop}, which is a proof automation tool, introduced the concept of
metavariable coupling, which is discussed further in
Section~\ref{sec:Background}. Metavariable coupling refers to the mutual
interference between goals in a proof, which complicates proof search.

More generally, a variety of \emph{hammers} have been developed for interactive
theorem provers.  These are generally based on automated reasoning (rather than
machine learning) techniques and include Isabelle's Sledgehammer
\cite{sledgehammer}, SMTCoq~\cite{smtcoq} and other hammers~\cite{coqhammer}
for Rocq, and various hammers~\cite{leansmt,leanauto,leanhammer} and the
Grind tactic in Lean.

Pantograph \cite{pantograph} affirmed the importance of the critical
observation that a proof may presented in a final form that differs significantly
from the structure of the proof as it was discovered and built a technical toolkit
in Lean to support these two different views of a proof. This is the
distinction between the \emph{search view} and \emph{presentation view}, which
we discuss more below. The concept of search view is related to the idea
of motivated proofs \cite{motivated-proofs}, where a proof is written in a way
which follows the author's trajectory of thinking rather than for easy
verification and mathematical beauty. We exploit this similarity to reduce the
search space of tactic agents. We use Pantograph to interact with Lean 4 in this
work, leveraging its expressive goal management and representation system.

MiniLang \cite{minilang} provides a reduced set of tactics for theorem proving
in Isabelle for language models. This set reduces the size of the action space
for a theorem proving agent. However, the size of an individual tactic can still
be arbitrarily large, as the \texttt{choose} tactic in MiniLang can provide an
arbitrarily-sized witness. In contrast, our atomic tactics always provide a
finite action space.


\section{Background}\label{sec:Background}
We provide some background on the setting of mechanical theorem proving in Lean.
Lean and Mathlib have large corpora of formally proved theorems, and these
theorems have been extensively used in training of machine learning agents. The
primary driver of theorem proving in Lean is the proof state, which
contains an \keyword{environment} of already-defined functions and lemmata, and a
set of goals to prove. Dispatching a tactic in this proof environment modifies the
set of goals. The mission of theorem proving is to discharge all goals in a
proof environment.

A newly discovered proof may be written in a different order than it was
conceived.
For example, in a common $\epsilon$-$\delta$ style convergence
proof, one is usually not able to come up with a correct $\delta$
before conducting further exploration. Consider proving that the sum of limits
$\hat a, \hat b$ of real functions $f$ and $g$ at $0$ is equal to
the limit of the sum $f(x) + g(x)$. To come up with this proof, we may opt
initially to
leave $\delta$ empty, i.e., $\delta := \square_\delta$. Then, we invoke the triangle inequality
on $|f(x) + g(x) - \hat a - \hat b| < \epsilon$ to find that we need to prove
$|f(x) - \hat a| + |g(x) - \hat b| < \epsilon$. This backward style of reasoning
is common in proof assistants such as Lean 4. We apply the definition of
convergence to arrive at $|f(x) - \hat a| < \square_f$ when $|x| < \delta_f$, where
$\delta_f$ is dependent on the yet uninstantiated $\square_f$, and likewise for
$\square_g$. Substituting, we find $\square_f + \square_g \leq \epsilon$, and
$\square_\delta \leq \min\{\delta_x, \delta_y\}$. At this point, we can solve for all the
uninstantiated variables using basic arithmetic.  A prover agent following a similar proof
trajectory does not have
to conjure complicated witnesses at the beginning and can instead rely on
incremental construction, finding witnesses at the end or along the way.

\subsection{Metavariables and Coupling}

Informally, a goal is a placeholder for an expression we
are trying to find of a particular type. These placeholders are called
\keyword{metavariables} in Lean.
Each metavariable also has a \keyword{context} consisting of a
set of \keyword{hypotheses}, free variables of particular types that
function as assumptions associated with the metavariable.
We write
$\Goal{g}[v_1,\dots,v_m]$ for a metavariable $\Goal{g}$ with an associated
context $v_1,\dots,v_m$.  Formally, a \keyword{goal} is a metavariable without an assigned value.
In general, however, metavariables can also have values assigned.
The type of a metavariable is an expression. For example,
a goal corresponding to the statement $p \land q$ is $\Goal{g} : p \land q$.
We write $\Context_{\Goal{g}}$ for the context of a goal
$\Goal{g}$ and $\Target_{\Goal{g}}$ for the type of $\Goal{g}$, also called the \keyword{target}.

A \keyword{tactic} is a Lean program that acts on one or more goals, assigns values to these
goals, and in the process may add new goals. A collection of goals is called
a \keyword{goal state} or just \keyword{state}.  The \texttt{constructor}
tactic for instance, when applied to $\Goal{g}$, generates an assignment for
$\Goal{g}$ and descendant goals $\Goal{h}_p$ and $\Goal{h}_q$.
\[
   \Goal{g} := \land.\mathsf{intro}(\Goal{h}_p, \Goal{h}_q), \qquad
   \begin{cases}
      \Goal{h}_p &: p \\
      \Goal{h}_q &: q
   \end{cases}
\]
In other words, one way to prove $p \land q$ is to prove $p$ and $q$
individually.

Metavariables are allowed to refer to each other. For instance, to prove
$\Goal{g} : \exists x \in \mathbb N,x - 2 = 0$, we can again use
the \texttt{constructor} tactic, generating two new goals:
\[
   \Goal{g} := \exists.\mathsf{intro}(\Goal{x}, \Goal{p}), \qquad
   \begin{cases}
      \Goal{x} &: \mathbb N \\
      \Goal{p} &: \Goal{x} - 2 = 0
   \end{cases}
\]
The first goal requires us to find a \keyword{witness} for the existential quantifier in
the original goal. Notice that $\Goal{p}$ mentions $\Goal{x}$ as part of its
type.  Whenever one goal refers to another, we follow the notation introduced by
Aesop \cite{aesop} and call them \keyword{coupled metavariables}. Because of
the possibility of coupling, proof search is not a simple tree search but is
rather a search on a directed acyclic graph (DAG). Given coupled goals
$\Goal{g}_1,\dots,\Goal{g}_n$, we must make a choice about which goal to make
progress on first, and the solution of one goal may impact the solution (or
indeed the provability) of a coupled goal.

For any goal $\Goal{g}$, we define the \keyword{cross-section} to be the number
of goals coupled to $\Goal{g}$ and its ancestors. Intuitively, the maximum
cross-section encountered during a proof can be seen as one measure
of the difficulty of a proof. As a general rule, tactics that produce lower
cross-sections are preferable to those that produce higher cross-sections. For
example, the problem of $\{\Goal{x}: \mathbb N, \Goal{p} : \Goal{x} - 2 = 0\}$
has a cross-section of 2, but $\{\Goal{h}_p : P, \Goal{h}_q : Q\}$, has a
cross-section of only $1$. The solutions of the goals $\Goal{h}_p$ and
$\Goal{h}_q$ do not interfere with each other.

In a given goal state, later goals provide cues for
solving earlier goals. For example, consider the goals
\begin{enumerate}
\item $\Goal{x} : \mathbb N$
\item $\Goal{1} : \Goal{x}\text{ is odd}$
\item $\Goal{2} : \Goal{x}\text{ is a perfect number}$
\end{enumerate}
It would be easy to close the goal $\Goal{x}$ by providing an arbitrary natural
number, but it could then be impossible to close the other two goals. It is sometimes
possible to close a coupled goal without explicit action. If we are proving
$\{\Goal{x}: \mathbb N, \Goal{p}: \Goal{x} \geq 2\}$, using the reflexivity
property of $\geq$ to prove $\Goal{p}$ would imply that $\Goal{x} := 2$. This
solves $\Goal{x}$ in passing. We will exploit this mechanism in
Section~\ref{sec:Methodology}.

\subsection{Proof Search}

Given the (mostly) tree structure of proofs, a natural way to search for a
proof is to use Monte-Carlo Tree Search \cite{mcts1987}.
There are two types of nodes in the search tree: goal nodes and goal state nodes.
A goal state node has a collection of goals. The decision that must be made at
a goal state node is \emph{which goal} to attempt to solve next.  We call this
the \keyword{guidance generation} problem.  Once a particular goal is selected, this
represents a transition from a goal state node to a goal node, in which a
single goal is present.  The decision that must be made at a goal node
is \emph{which tactic} to apply to attempt to solve the goal.  We call this
the \keyword{tactic generation} problem.  Applying a tactic to a goal
may either solve the goal, which closes the current search branch, or it may
generate new goals to solve, representing a transition from a goal node to
another goal state node.

\subsection{Proof Views}

\begin{figure}
	\centering
	\begin{tikzpicture}
		\node[draw,rounded rectangle] (pv) at (0:3) {Presentation View};
		\node[draw,rounded rectangle] (sv) at (120:3) {Search View};
		\node[draw,rounded rectangle] (kv) at (240:3) {Kernel View};
		\draw
			(pv.75)   edge[auto=right,->] node[sloped] {Articulation} (sv.290)
			(sv.250)  edge[auto=left, ->] node[sloped] {\keyword{Transposition}} (pv.115)
			(kv.45)   edge[auto=right,->] node[sloped] {Delaboration} (pv.-75)
			(pv.-115) edge[auto=left, ->] node[sloped] {Elaboration} (kv.75)

			(kv.130)  edge[auto=left,->] node[sloped] {Essentialization} (sv.230)
			(sv.210)  edge[auto=right,->] node[sloped] {Contextualization} (kv.150)
			;
	\end{tikzpicture}
	\caption{The Three Views of a Proof and their relations. We discuss
	Transposition in Section~\ref{sec:transpose} and Essentialization and
	Contextualization in Section~\ref{sec:Graph}.  We do
	not dicsuss Articulation.}
	\label{fig:three views}
\end{figure}

In Lean, there are many different ways to prove a theorem. Any theorem proving
agent or human operator interacting with Lean must pick a method for receiving
goals from Lean and sending tactics into Lean. For language models and people,
the default choice is strings, provided by Lean's \emph{delaborator}. Conversely,
human and machine written tactics produce Lean expressions by executing arbirary
Lean code.

We adopt the terminology of Pantograph~\cite{pantograph} in distinguishing between three
broad categories of proof representation styles (see Figure~\ref{fig:three views}). Internally, Lean stores proofs and goals in
the \keyword{kernel view}, as a set of assigned and unassigned metavariables, and delaborates them to
the \keyword{presentation view} on demand, which is a string-based
representation that prioritizes ease of understanding and verification.
Presentation view is concise and typically has no coupling.  However, a proof
written for presentation often includes mysterious and unintuitive jumps,
especially when introducing new expressions. This motivates the concept of a
proof in \keyword{search view}, a representation corresponding to the proof tree
structure mentioned above. A search view proof tracks the path of an agent
searching for a proof and may contain coupling and backtracking. More
importantly, there may be information present in the kernel and presentation
views that is elided in the search view, since it is not relevant for a
particular goal. For
example, for goals $\Goal{g}: P, \Goal{h}: 2 = \Goal{x}$, the solution of
$\Goal{h}$ has no impact on the solution of $\Goal{g}$ and therefore is
irrelevant to $\Goal{g}$ in the search view. In Section~\ref{sec:Graph}, we
describe the process of converting a goal into search view.
The $\epsilon$-$\delta$ convergence proof mentioned above also illusrates the difference between
presentation and search views.




\section{Atomic Tactics and Atomization}\label{sec:Methodology}
\begin{table}[t]
	\centering
	\begin{tabular}{l|l|l}
		\hline
		Atomic & Constructors & Meaning \\
		\hline
		Invalid & \texttt{.bvar} & Bound Variable \\
		Invalid & \texttt{.mvar} & Metavariable \\
		\texttt{inhabit} & \texttt{.sort} & Type Sort \\
		\texttt{intro} & $\func x.y$ & Lambda function \\
		\texttt{exact}/\texttt{apply} & \texttt{.fvar } & Free variable \\
		\texttt{inhabit} & \texttt{.lit} & Literal \\
		\texttt{exact}/\texttt{apply} & \texttt{.const} & Constant \\
		\texttt{pi} & $\forall x.y$ & Function Signature \\
		\texttt{tailArg} & \texttt{.app} & Function Application \\
		Unfold & \texttt{.letE} & Let-In binder \\
		Unfold & \texttt{.mdata} & Expression Metadata \\
		\texttt{apply}/\texttt{cases} & \texttt{.proj} & Projection \\
		\hline
	\end{tabular}
        \vspace*{.3cm}
	\caption{Correspondence between atomic tactics and elementary expression
	constructors. The constructors labeled ``Invalid'' cannot appear at the
        top-level of a proof term. The constructors labeled ``Unfold'' can be
        transformed into another case.}
        \vspace*{-.5cm}
	\label{table:Atomic tactic to expression}
\end{table}

A tactic generation agent faces the problem of
the immense number of possible
tactics available for constructing proofs. In this section, we propose one
possible mitigation for this difficulty by introducing the notion of an atomic
set of tactics and an algorithm called atomization for generating atomic proofs
from arbitrary proofs.

\subsection{Atomic Tactics}
A set of tactics is \keyword{atomic} if it has two key features. First of all,
it must only permit a finite number of actions (for parametric tactics, we
require the number of actions to be finite even when considering all possible
values of the parameters). In contrast, tactics used in standard Lean proofs
are often non-atomic in this sense. For instance, the \lstinline{simp} tactic
takes an arbitrarily long parameter list, and the \lstinline{conv} tactic
environment allows for arbitrarily long navigation sequences.  The
\lstinline{use} tactic in Lean is especially non-atomic, since the witness
expression could be arbitrarily long. Atomic tactics are designed to avoid such
situations. As a result, an agent that is trained to pick from a set of atomic
tactics need only consider a finite number of possibilities, greatly
simplifying the task. The other key feature is that a set of atomic tactics
should be complete in the sense that it should be sufficiently expressive to
prove any valid Lean theorem.

In Figures~\ref{fig:Atomic Tactics from Lean} and \ref{fig:Atomic Tactics Synthetic}, we present a concrete set of atomic tactics.  The first feature is
ensured by construction---each atomic tactic has only a finite number of
parameter possibilities. This means in particular that no tactic takes an
arbitrary number or string, as that would make its action space infinite.
For example, the
\lstinline{inhabit} tactic does not take an argument---instead
it generates some default value for an inhabited type. The \lstinline{apply}
tactic takes exactly one lemma argument, and in any proof state, only a finite
number of lemmas are available in the environment. Note, however, that it is still
possible for a proof using atomic tactics to assign arbitrary string or number
literal values to a goal.  This can happen, for example, by applying a lemma
that assigns a specific solution value.

To ensure completeness, we include at least one atomic tactic
corresponding to each top-level expression constructor that could appear in a
proof term in Lean.  Table~\ref{table:Atomic
tactic to expression} shows the different constructors that can appear
in a proof term and the corresponding atomic tactic.  Those marked ``Invalid''
cannot occur as the top-level constructor of a proof term, and those marked
``Unfold'' can be transformed into a equivalent expression which has a different
constructor.

There is one more desirable feature for a set of atomic tactics. They should be
chosen to minimize the cross-sections of proofs. Proofs with large
cross-sections pose a greater challenge for a theorem proving agent, as it must
somehow choose which of many coupled goals to try to make progress on.


With these objectives in mind, we now explain the chosen set of atomic tactics.%
\footnote{The full list of tactics can be found in Appendix~\ref{sec:Atomic Tactics}.}
We rely on the following properties of our atomic tactics.  These properties
are necessary for the atomization and agent training steps we describe next.
\begin{enumerate}
\item Invariance: No tactic ever changes the value of an assigned
metavariable, i.e., once a goal is assigned, its assignment never changes without
	backtracking.
\item Completeness: If a tactic removes a goal, it must assign the goal.
\item Progress: It is not possible for a tactic to make no progress, i.e.,
	a tactic never produces a goal state containing only the goal it was
	applied to.
\item Determinism: The atomic tactics yield the same result for the same input
	goal.
\end{enumerate}


\subsection{Transposing Atomization}\label{sec:transpose}

Recall that our motivation for creating a set of atomic tactics is to ease the
burden on an agent tasked with solving the tactic generation problem.  But if we
are to train such an agent, we need training data in the form of successful
\emph{search view} proofs made up of \emph{atomic tactics}.  We next show how to
obtain such proofs via a novel \keyword{transposing atomization} algorithm. This
algorithm has two important goals.  The first goal is to convert arbitrarily
large or complex tactics and expressions into atomic tactics.  We call this
\keyword{atomization}.  The second goal is to convert presentation view proofs
into search view proofs.  We call this \keyword{transposition}. The algorithm
takes as input the ground truth proof term of a theorem, $e_{\mathsf{root}}$, and
outputs a sequence of goal tactic pairs indicating the actions on individual
goals.

\newcommand{\proofexpr}{\ensuremath{\mathit{proofexpr}}\xspace}
\newcommand{\atomizestep}{\ensuremath{\mathit{atomize\_step}}\xspace}

The starting point for atomization is a ground-truth proof term $\proofexpr : T$
which proves $T$, also called a \keyword{solution} for $T$.  We use $\proofexpr$
to guide the construction of a new proof for $T$ constructed using only
atomic tactics.  Pseudocode for the algorithm is shown in Listing 1.
We start by creating a goal for $T$, and
inserting it and its solution into a list of pending goals.  Then, for each pair in the
list, we apply the $\atomizestep$ function, which
tries to find an atomic tactic that can be used on the goal.  If successful, it
returns the tactic, a set of new goals
produced by the tactic when applied to the goal, and,
crucially, corresponding solutions for each new goal. The new goals and
solutions are then inserted back into the list, and the process repeats until
the pending list is empty.  If no atomic tactic can be found, the goal is
deferred, hopefully becoming solvable later.

The atomize function returns a list of pairs, each containing a goal
and the corresponding atomic tactic computed for that goal.  This gives us a recipe for proving $T$.
We simply start with the goal $\Goal{g} : T$, look up which atomic tactic is paired with
$\Goal{g}$, and apply that tactic to get new goals.  We then pick one of the new goals
$\Goal{g'}$, look up which atomic tactic is paired with $\Goal{g'}$, and continue until no
goals are left.

Atomization is easy in the vast majority of cases. For
example, if the goal: $\Goal{g} : \forall (x : X), Y$ has the solution:
$\func x \mapsto y$, the natural step is to associate this goal with the
\texttt{intro} tactic.  Applying the tactic to the goal produces a new goal
$\Goal{g'} : Y$ whose corresponding solution is just $y$.  We do not have space to go into
details for every atomization step, but the code is available with all of the
details.\footnote{Code will be released after the anonymous review period.}

By stretching out expressions into many tactics, we can generate a large amount
of training data from Lean's standard library and Mathlib. Nazrin is designed in
conjunction with a set of atomic tactics using 3 key principles:
\begin{enumerate}
\item Minimization of Cross-section: Atomization should not generate a large
	amount of coupled goals, which pose a immense challenge to guidance
	generation discussed in Section~\ref{sec:Graph}.
\item Completeness: Atomic tactics should be able to repeat proofs written in
	non-atomic tactics.
\item Non-Volatility: The solution must be replay-able.
\end{enumerate}

\begin{figure*}
\centering
\begin{subfigure}{.5\textwidth}
	\centering
	\includegraphics[width=.8\linewidth]{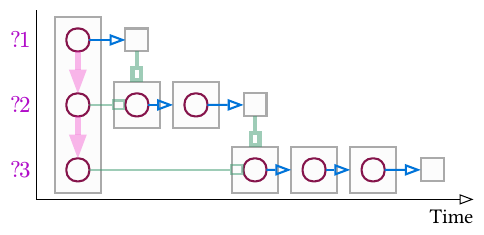}
	\caption{Tactic dispatch order of a proof}
\end{subfigure}%
\begin{subfigure}{.5\textwidth}
	\centering
	\includegraphics[width=.8\linewidth]{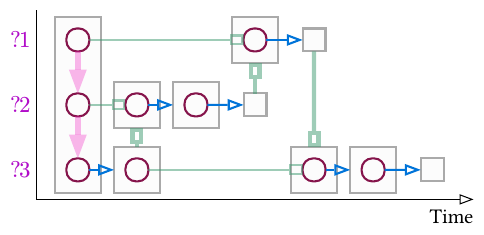}
	\caption{Tactic dispatch order of a state-level transposed proof}
\end{subfigure}
\caption{Blue arrows represent tactic applications. State-level transposition changes the dispatch order of tactics, but does not change the ordering of tactics on a particular goal.}
\label{fig:state-level transposition}
\end{figure*}

There are a few more features of the algorithm worth explaining.  The first is
the question of what order to process the goals in.
Define a goal $\Goal{h}$ to be an \keyword{immediate predecessor} of goal
$\Goal{g}$ (and $\Goal{g}$ is the \keyword{immediate successor} of $\Goal{h}$),
denoted $\Goal{h} \prec \Goal{g}$, if an expression in $\Goal{g}$'s target or
context contains $\Goal{h}$.\footnote{As a subtle exception to this rule, we
don't include cases where $\Goal{h}$ has any parent expression whose type is a
$\mathsf{Prop}$.}  $\Goal{g'}$ is a \keyword{successor} to $\Goal{g}$ if there
is a chain of successors $\Goal{g} \prec \cdots \prec \Goal{g'}$.  When we have
a choice of goals to atomize, we preferentially choose successor goals over
predecessor goals.
The solution of these goals
hopefully sheds light on the solutions of their predecessors or solves the
predecessors in passing. We call this process \keyword{state-level
transposition}.  A visualization is in Figure~\ref{fig:state-level transposition}.

\begin{lstlisting}[caption={Nazrin's Transposing Atomization},label=list:8-6,captionpos=t,float,abovecaptionskip=-\medskipamount]
function atomize(proofexpr)
   goal := Goal(infer_type(proofexpr))
   pending := [(goal, proofexpr)]
   while (goal, solution) := pending.pop()
      if let Some((tactic, goals, solutions))
            := atomize_step(goal, solution)
         yield (goal, tactic)

         descendants := order_predecessor(goals, solutions)
         for (g, s) in (goals, solutions)
             pending.push(g, s)
      else
         yield (goal, :defer)
      end
   end
end

# If we can create an atomic step on this goal, return a tactic and
# solutions to descendant goals.
function atomize_step(goal, solution)
   if (tactic, solutions) := try_semigrade(goal, solution)
      return (tactic, solutions)
   else if (tactic, solutions) := try_holograde(goal, solution)
      return (tactic, solutions)
   else
      return :defer
   end
end
\end{lstlisting}

In the transposing atomization algorithm, we maintain an invariant that every
goal will eventually acquire a unique assignment. We define the
\keyword{closure} $\Goal{\Closure{g}}$ of $\Goal{g}$ to be the goal formed by
instantiating all predecessors of $\Goal{g}$.

During atomization, each generated goal $\Goal{g}$ is paired with an expression
$e_{\Goal{g}}$, called the \keyword{solution}, with the invariant
$\Context_{\Goal{\Closure{g}}} \vdash e_{\Goal{g}} : \Target_{\Goal{\Closure{g}}}$
maintained at all times. i.e. the solution solves the completed version of goal.
We dispatch the next tactic based on $e_{\Goal{g}}$, and if no tactics can
dispatch on $\Goal{g}$, we defer to another goal. In this case $\Goal{g}$
becomes \emph{dormant}.

During atomization, Lean's usually helpful type unification algorithm works
against us. If used naively, type unification algorithm may assign incorrect solutions to
goals and back the atomization algorithm into an impossible corner. To avoid this problem, we
rely in such cases on the synthetic tactics \lstinline{tailArg} and
\lstinline{motivatedApply}, which do not do type unification. The cost
of running these tactics is that they generate large cross-sections.

\begin{figure*}
	\centering
	\includegraphics[width=.8\textwidth]{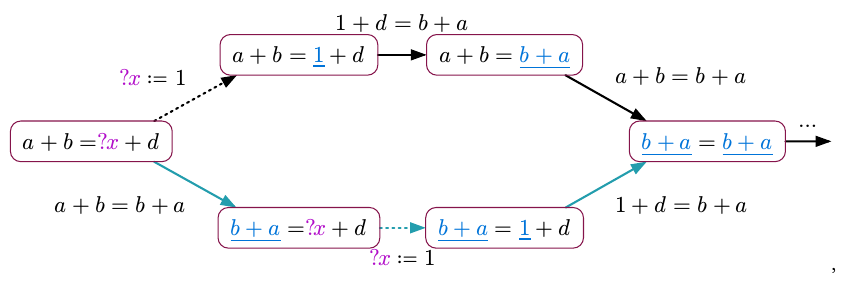}
	\caption{Goal-level transposition switches orders of tactics on the same
	goal. Although $\Goal{x}$ is not assigned, we can dispatch a rewrite tactic
	if we rearrange the order of semigrade tactics.}
	\label{fig:rewrite semigrade}
\end{figure*}

Within the process of proving a goal, the execution order of tactics can often be rearranged. We call
this \keyword{goal-level transposition} and leverage it to avoid dead ends that
our limited set of tactics would otherwise encounter.
We divide the set of atomic tactics into 7 categories, based on
their \emph{area of effect}, which is the smallest subexpression a tactic modifies:
a \keyword{holograde} tactic acts on an entire free variable or
target; a \keyword{semigrade} tactic acts on parts of a free variable or target;
a \keyword{prograde} tactic acts on a free variable; and a \keyword{retrograde}
tactic acts on the target. Evidently, two semigrade tactics can be dispatched in
arbitrary order as long as their areas of effect do not collide. For example,
rewriting $a = b$ and $c = d$ can dispatch in any order on $a = c$. Finally, a
\keyword{bigrade} tactic acts on both the target and a free variable. In
Figure~\ref{fig:rewrite semigrade}, we show an example where the execution order
of two semi-retrograde rewrite tactics are swapped due to non-conflicting acting
areas. Since the coupled goal $\Goal{x}$ has not been instantiated at the
beginning, this goal-level transposition allows 2 tactics to dispatch instead of
1. We use goal-level transposition to dispatch as many tactics as possible on a
goal before deferring to a predecessor goal. A prograde and a retrograde tactic
can likewise swap without interfering with the validity of a proof. A
\keyword{terminal} tactic closes a goal.

\begin{table}
	\centering
	\begin{tabular}{l|l}
		\hline
		Category & Examples \\
		\hline
		Terminal & \texttt{decide} \\
		Holo-prograde & \texttt{cases} \\
		Holo-retrograde & \texttt{apply}, \texttt{intro} \\
		Holo-bigrade & \texttt{revert}, \texttt{induction} \\
		Semi-prograde & \texttt{rewrite} (on free variable) \\
		Semi-retrograde & \texttt{rewrite} (on target) \\
		Semi-bigrade & \texttt{unfold} \\
		\hline
	\end{tabular}
        \vspace*{.3cm}
	\caption{Examples of each grade of tactics.}
        \vspace*{-.5cm}
	\label{table:Example of tactic grades}
\end{table}

Note that the extra goals generated by a tactic are not considered. For semigrade tactics such as \texttt{rewrite}, the main goal (with
its target/local assumption modified by the rewrite) is semi-retrograde or
semi-prograde, but the tactic generates extra goals on the side. Since the
further processing of these goals is not hindered by the main goal,
they do not contribute to the grade system. An example is a rewrite lemma
that is gated by some condition $h: p(a) \to f(a) = b$. Applying this lemma
to $\vdash \mask{f(a)} = f(c)$ generates two goals, $\vdash \Goal{1} : b = f(c)$
and $\vdash \Goal{2} : p(a)$. We can dispatch another semigrade tactic
on $f(a) = \mask{f(c)}$ without interfering with the operation of this lemma,
but any holo-retrograde tactic would not be able to dispatch ahead of this
rewrite due to conflict with the rewrite.

Atomization also provides a metric for the difficulty of a proof. A
proof whose atomization trace contains large cross sections requires the
simultaneous consideration of many factors. A proof whose atomization is long
requires more steps. The benefit of this measurement compared to number of
lines or number of tactics is that 
proofs with few but arbitrarily long tactics cannot hide their difficulty by
simply having many tactic parameters. Note that our transposing atomization algorithm
could be implemented in other proof assistants, including Rocq and Isabelle.


\section{ExprGraph and Graph Neural Networks}\label{sec:Graph}
\begin{figure*}[t]
	\centering
	\begin{tikzpicture}[
		xscale=2, yscale=0.8,
		Mora/.style={draw},
		Begin/.style={rounded rectangle},
		Middle/.style={rectangle},
		Premise/.style={fill=ForestGreen!50,Middle},
		Terminal/.style={chamfered rectangle},
		Free/.style={fill=blue!30,Middle},
		Locus/.style={fill=Dandelion!30,Middle},
		Label/.style={midway,fill=white},
	]
		\node[Mora,Begin] (kind) at (0,0) {Kind};
		\node[Mora,Terminal] (exit) at (6,0) {Exit};
		\draw[->] (kind) -- (exit) node[midway,fill=white] {\texttt{intro}, abandon};
		\node[Mora,Premise] (apply) at (1,-1) {Apply};
		\node[Mora,Premise] (induction) at (1,-2) {Induction};
		\node[Mora,Free] (induction major) at (2,-2) {Major};
		\node[Mora,Premise] (rewrite) at (1,-3) {Rewrite};
		\node[Mora,Locus] (rewrite locus) at (2,-3) {Locus};
		\node[Mora] (rewrite dir) at (3,-3) {Direction};
		\draw[->] (kind) -- (kind |- apply) -- (apply);
		\draw[->] (apply) -- (apply -| exit) node[Label] {\texttt{apply}} -- (exit);
		\draw[->] (kind) -- (kind |- induction) -- (induction);
		\draw[->] (induction) -- (induction major);
		\draw[->] (induction major) -- node[Label] {\texttt{induction}} (induction major -| exit) -- (exit);
		\draw[->] (kind) -- (kind |- rewrite) -- (rewrite);
		\draw[->] (rewrite) -- (rewrite locus);
		\draw[->] (rewrite locus) -- (rewrite dir);
		\draw[->] (rewrite dir) -- node[Label] {\texttt{rewrite}}  (rewrite dir -| exit) -- (exit);
	\end{tikzpicture}
	\caption{Neural Probabilistic Automaton for Tactic Generation.
\textcolor{ForestGreen}{Green states} perform premise selection,
\textcolor{blue}{Blue states} perform free variable selection.
\textcolor{Dandelion}{Yellow states} perform locus selection. White states perform
fixed-length categorical selection. Tactics without parameters are dispatched
	directly by the Kind state and hence do not show up as states of the NPA.}
	\label{fig:NPA}
\end{figure*}

We expect atomic tactics and atomization to be useful for a variety of
applications, but one main motivation is to provide training data for a theorem
proving agent. In this section, we explain the design of such an agent based on
Graph Neural Networks (GNNs). We selected this architecture since a graph can
compactly represent many symmetries inherent in mathematics, and we hypothesize
that GNNs will thus be an efficient mechanism for theorem proving.

\subsection{ExprGraphs}
The first problem we address is how to represent Lean expressions and goals
in a format which is suitable for consumption for a machine learning agent, a
GNN in our case.
We call the process of converting an expression to a graph \keyword{Essentialization}. In the process of essentialization,
we carefully discard irrelevant information in the solution
of the current goal. For example, an expression
$\mathsf{List}.\mathsf{head}\,a\,h$ queries the first element of a list, and
this query requires a certificate $h$ that the list is non-empty. However, exactly how
$h$ was generated is irrelevant to any goal containing this expression. For this
reason, $h$ should be removed from the expression in question.

Essentialization erases some technical distinctions that are irrelevant to
searches. For example, consider $\Goal{g}$ and $\Goal{g'}$ under two different
contextes:
\[
	\begin{cases}
		&\vdash \Goal{g} : f(\Goal{1}) \\
		&\Goal{1} := \func x \mapsto \Goal{2}[x]
	\end{cases} \qquad
	\begin{cases}
		&\vdash \Goal{g'} : f(\func x \mapsto \Goal{2}[x])
	\end{cases}
\]
In the search view, $\Goal{g}$ and $\Goal{g'}$ are identical---the same sequence of tactics that proves $\Goal{g}$ will lead to the
proof of $\Goal{g'}$ and vice versa---but they are
different in the kernel view since the former involves an additional
metavariable $\Goal{1}$.
We say $\Goal{g}$ and $\Goal{g'}$ are \keyword{search-view equivalent}. If
goal $\Goal{h}$ is not coupled to $\Goal{g}$, then $\Goal{h}$ would not
participate in the ExprGraph for $\Goal{g}$. This is in contrast to Lean's
infoview which displays all unsolved goals.

\begin{figure}
	\centering
	\includegraphics[width=.3\textwidth]{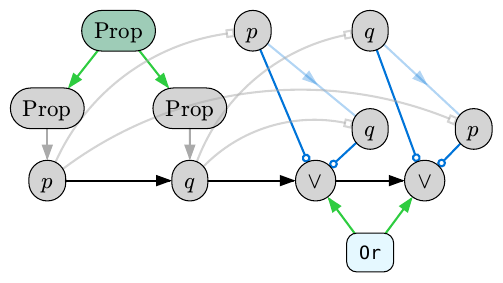}
	\caption{ExprGraph of $\forall (p\,q: \Prop), p \lor q \to q \lor p$.
	Notice that every subexpression corresponds to a locus (grey vertex) in this
	graph.}
\label{fig:exprgraph}
\end{figure}

Each essentialized expression and goal is represented as an \keyword{ExprGraph}, a
heterogeneous graph whose nodes are the syntactic elements and whose edges
represent relationships between them.
This allows graph message passing to transmit information between related parts
of the graph and between coupled goals.
A simple example is shown in Figure~\ref{fig:exprgraph}.
An ExprGraph $G(e)$ of an expression $e$ is designed to have several desirable properties:



\begin{enumerate}
\item \keyword{Symmetry}: If two expressions are \textalpha-equivalent or search-view
	equivalent, they have the same ExprGraph. This erases immaterial distinctions in the
	kernel view. For example, $\int f(x)\,\mathrm{d}x$ and
	$\int f(y)\,\mathrm{d}y$ have identical meanings and produce the
	same ExprGraph.
\item \keyword{Self-Similarity}: If $e_1$ is a subexpression of $e_2$, then
	$G(e_1)$ is a subgraph of $G(e_2)$.
\item \keyword{Locus Conservation}: Every rewritable subexpression
	(\keyword{locus}) of $e$ corresponds uniquely to a vertex in $G(e)$.
\item \keyword{Condensation}: All references to the same constant, sort, or
	literal in an expression are connected with a single shared vertex.
\end{enumerate}

\begin{figure}
\centering
\begin{subfigure}{.4\textwidth}
	\centering
	\includegraphics[width=.6\textwidth]{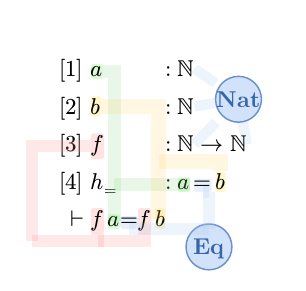}
	\caption{Essentialization erases variable names}
\end{subfigure}

\begin{subfigure}{.4\textwidth}
	\centering
	\includegraphics[width=.6\textwidth]{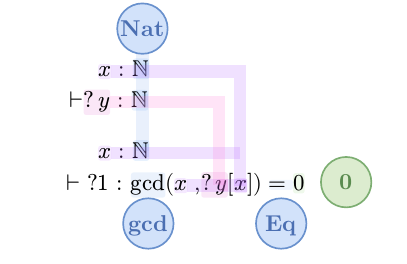}
	\caption{Essentialization cross-references variables in coupled goals}
\end{subfigure}
\caption{We visualize key aspects of goals and the corresponding ExprGraph.
Colored lines highlight edges which replaces variable names. Circles represent
nodes that consolidate all references to a constant.}
	\label{fig:Essentialization of Goals}
\end{figure}

\noindent
Figure~\ref{fig:Essentialization of Goals} visualizes essential aspects of ExprGraphs.

\subsection{Tactic Generation from Graph Neural Networks}

The set of atomic tactics effectively makes the proof search action space
finite. This finiteness reopens many toolboxes from classical reinforcement
learning and makes the model more diagnosable. If none of our
tactics had parameters, our model could simply output a probability
distribution. The generation of tactics with parameters is slightly more
complicated. Since a graph neural network is not a generative model, it cannot
easily generate parameterized tactics. However, armed with atomic tactics, we can use
the GNN to condition the probabilities inside a non-deterministic
automaton. We call this architecture a \keyword{neural probabilistic automaton}
(\keyword{NPA}).
The transition probability from one state to the next
is defined by the neural network with a categorical distributional output. An NPA
is a highly specialized heterogeneous autoregressive model that can create
finite length tactics.

The first state (called ``Kind'') of the NPA selects which tactic to use. Then,
depending on the tactic, additional specialized GNN heads generate arguments to
the tactic in descendant states. For instance, the Kind state can pick the
\lstinline{intro} tactic, which leads to the termination of the NPA. On the
other hand, if the Kind state picks the \lstinline{apply} tactic, we then
consult the apply state to pick a constant. This is a retrieval problem where
the query is an embedding vector, the keys are embedding vectors for all usable
constants, and the values are Lean constant names. A subset of our NPA is shown
in Figure~\ref{fig:NPA}.

The generated tactic has to be \keyword{contextualized} into a form Lean can
understand. This includes translating locus indices into subexpression positions
and translating free variable indices to free variable names. The contextualized
tactic then dispatches in Lean.

We construct the GNN out of 4 components: an embedding layer which contains
embeddings of individual constants, a core equivariant GNN which is responsible
for in-depth understanding of an expression, a fixed-point invariant GNN which
generates embeddings from unseen constants, and individual tactic heads to
control probability distributions on each state. The fixed-point invariant GNN
is used to generate embeddings for unseen new functions and theorems, which
allows the GNN to operate on symbols that are not in its training set.


We use an embedding dimension of $32$, a core GNN using 5 attention convolution
layers of 4 heads each, and the gelu activation function \cite{gelu}. It has a
total of 1.5 million parameters when running on standard library and
11 million parameters on Mathlib, most of which are in the embedding layer. We
execute intro and terminal tactics with mechanical assistance: these tactics are
automatically tried before consulting the graph neural network.

The Nazrin Prover is small enough to train and evaluate on a consumer-grade CPU-only
machine.
In comparison to the seconds-level tactic generation speed of language models,
Nazrin and other GNN-based models can generate thousands of tactics per minute,
enabling highly parallelized proof searches. Since Nazrin is trained from
scratch on atomization data, we can ensure the absence of data leakage from the
training set to the test set. We train Nazrin Prover with exclusively supervised
learning data and no online data. This is in contrast to Q-function learning
where the model learns to estimate the reward of actions.


In Nazrin, we use a mechanical form of guidance generation known as
\keyword{rainbow guidance}: If $\Goal{g} \prec \Goal{h}$, then the agent assigns
a higher priority to $\Goal{h}$. This encourages an agent to first explore
successors rather than predecessors. If there is a tie, we use the ordering of
goals within a state as the tie-breaker. On any goal $\Goal{g}$, the tactic
agent can emit a special \emph{abandon} action, indicating that no more progress
can be made. This abandon semantics is used for two cases: If the search agent
believes it is futile to continue searching on a goal, and if the current goal's
predecessors need further progress and information before any tactic can make
progress on a goal.

\section{Evaluation}\label{sec:Evaluation}
We evaluate Nazrin using Lean v4.25.2 and Julia v1.12.4 for running Graph Neural
Networks. We first atomized 170180 user-defined theorems in Lean's standard library
and Mathlib.

Atomized theorems generally have low cross-sections as shown in Figure~\ref{fig:Atomization
dataset}.  The success rate of atomization is about $58\%$. We set a limit of
3000 maximum steps due to computation time limits. The atomization algorithm can
fail if this maximum is exceeded or if the current heuristics fail to find a
transposition.

\begin{figure*}[t]
	\centering
	\begin{subfigure}{.5\textwidth}
		\centering
		\includesvg[width=.9\linewidth]{diagrams/result/cross-section-stdlib.svg}
		\caption{Standard Library}
	\end{subfigure}%
	\begin{subfigure}{.5\textwidth}
		\centering
		\includesvg[width=.9\linewidth]{diagrams/result/cross-section-mathlib.svg}
		\caption{Mathlib}
	\end{subfigure}
   \caption{Maximum cross-section and proof length in atomized proofs.}
	\label{fig:Atomization dataset}
\end{figure*}

\begin{figure*}[t]
   \centering
	\begin{subfigure}{.5\textwidth}
		\centering
		\includesvg[width=.9\linewidth]{diagrams/result/search-aesop.svg}
		\caption{Runtime/Completion Comparison with Aesop}
	\end{subfigure}%
	\begin{subfigure}{.5\textwidth}
		\centering
		\includesvg[width=.9\linewidth]{diagrams/result/search-grind.svg}
		\caption{Runtime/Completion Comparison with Grind}
	\end{subfigure}
   \caption{Comparison of Nazrin Prover with Aesop and Grind on a sample of
	slice 4 of Mathlib.  Both provers are allotted 15 seconds. The bottom-right data points
	represent theorems that Nazrin Prover can prove but aesop/grind
	cannot.}
	\label{fig:Compare with automation}
\end{figure*}

We then sorted the atomized theorems topologically to ensure that all
dependencies of a theorem precede the theorem. This ensures the absence of
information leakage during training. Our evaluation setup mimics a real use
case, where a user is trying to develop a new formalized theory based on
Mathlib. We divided the theorems in the standard library into 2 slices and
those in Mathlib into 10 slices. We measure the generalizability of Nazrin
Prover by training it on slice $i$ and evaluating it on slice $i+1$. The metric
for evaluation is the rate of successful proofs within a 15-second time
limit. With our atomized dataset, we trained Nazrin Prover on the standard
library slice 1 for 100 epochs. Each segment has about 10000 theorems. After
training on stdlib slice 1, we evaluate Nazrin Prover on stdlib slice 2,
achieving an accuracy of $57\%$ percent on a random subset
(Appendix~\ref{sec:Evaluation Result on Stdlib}).  We then trained Nazrin Prover
on Mathlib slice 3, and mesaured its proof completion rate on slice 4. This
reached $34\%$. We compare with Aesop \cite{aesop} and Grind \cite{lean4}
automation tactics in Figure~\ref{fig:Compare with automation}.\footnote{For
Aesop and Nazrin, we use perfect premise selection, both to ensure fairness and
to prevent Aesop from proving a theorem using itself.  Nazrin does not have to
operate with perfect premise selection in production.  There is no easy way to
do this for Grind, so Nazrin is at a disadvantage in this experiment.}  Observe
that Nazrin Prover is capable of proving theorems that other automation tactics
cannot discharge.

\section{Conclusion}\label{sec:Conclusion}
We introduce atomic tactics which provide
a finite action space for prover agents. We describe a transposing
atomization algorithm, which converts existing proofs into atomized proofs. We
introduce Nazrin Prover, a GNN-based high-throughput and high-performance
theorem proving agent. We evaluate the agent on generalization tasks on Lean's
standard library and Mathlib, demonstrating complementary capabilities with
other proof automation tactics.


Atomization does not yet work for all theorems. Future work could improve the
coverage for atomization in Mathlib. The atomization algorithm sometimes generates
proofs of high cross sections. This places a burden on the prover agent to find
one goal out of many to make progress on. Future work could focus on obtaining
lower cross-sections.


The architecture of Nazrin Prover has not been carefully tuned to maximize
performance. Future work could further refine its architecture and boost
performance. In particular, a neural network could replace the existing
mechanically generated rainbow guidance. Moreover, in tactics with multiple
parameters, the later
parameters are not conditioned on the former parameters. We made this
simplification to accelerate training and inference. This may make it difficult
for the NPA to generate such tactics.
Our results with Nazrin Prover could also likely be improved with more training.

The GNN does not process numbers and strings. In future work, another model or
a mechanical method could take over to resolve these problems. We work around
this issue by assigning goals in passing using Lean's type unification system.

%


\bibliography{references.bib}

@proceedings{lean4,
	author="de Moura, Leonardo and Ullrich, Sebastian",
	editor="Platzer, Andr{\'e}
	and Sutcliffe, Geoff",
	title="The Lean 4 Theorem Prover and Programming Language",
	booktitle="Automated Deduction -- CADE 28",
	year="2021",
	publisher="Springer International Publishing",
	address="Cham",
	pages="625--635",
	isbn="978-3-030-79876-5"
}

@misc{pantograph,
	title={Pantograph: A Machine-to-Machine Interaction Interface for Advanced Theorem Proving, High Level Reasoning, and Data Extraction in Lean 4},
	author={Leni Aniva and Chuyue Sun and Brando Miranda and Clark Barrett and Sanmi Koyejo},
	year={2025},
	eprint={2410.16429},
	archivePrefix={arXiv},
	primaryClass={cs.LO},
	url={https://arxiv.org/abs/2410.16429},
}

@inproceedings{aesop,
	author = {Limperg, Jannis and From, Asta Halkj\ae{}r},
	title = {Aesop: White-Box Best-First Proof Search for Lean},
	year = {2023},
	isbn = {9798400700262},
	publisher = {Association for Computing Machinery},
	address = {New York, NY, USA},
	url = {https://doi.org/10.1145/3573105.3575671},
	doi = {10.1145/3573105.3575671},
	booktitle = {Proceedings of the 12th ACM SIGPLAN International Conference on Certified Programs and Proofs},
	pages = {253–266},
	numpages = {14},
	location = {Boston, MA, USA},
	series = {CPP 2023}
}

@article{mathlib,
	title     = {The Lean mathematical library},
	author    = {mathlib},
	journal   = {CoRR},
	volume    = {abs/1910.09336},
	year      = {2019},
	url       = {http://arxiv.org/abs/1910.09336},
	eprinttype = {arXiv},
	eprint    = {1910.09336},
	bibsource = {dblp computer science bibliography, https://dblp.org}
}

@inproceedings{mcts1987,
	author = {Abramson, Bruce and Korf, Richard},
	year = {1987},
	month = {01},
	pages = {90-94},
	title = {A Model of Two-Player Evaluation Functions.}
}

@misc{hypertree,
      title={HyperTree Proof Search for Neural Theorem Proving},
      author={Guillaume Lample and Marie-Anne Lachaux and Thibaut Lavril and Xavier Martinet and Amaury Hayat and Gabriel Ebner and Aurélien Rodriguez and Timothée Lacroix},
      year={2022},
      eprint={2205.11491},
      archivePrefix={arXiv},
      primaryClass={cs.CL}
}

@inproceedings{wang2023dt,
	address={Toronto, Canada},
	title={DT-Solver: Automated Theorem Proving with Dynamic-Tree Sampling Guided by Proof-level Value Function},
	url={https://aclanthology.org/2023.acl-long.706},
	DOI={10.18653/v1/2023.acl-long.706},
	abstractNote={Recent advances in neural theorem-proving resort to large language models and tree searches. When proving a theorem, a language model advises single-step actions based on the current proving state and the tree search finds a sequence of correct steps using actions given by the language model. However, prior works often conduct constant computation efforts for each proving state while ignoring that the hard states often need more exploration than easy states. Moreover, they evaluate and guide the proof search solely depending on the current proof state instead of considering the whole proof trajectory as human reasoning does. Here, to accommodate general theorems, we propose a novel Dynamic-Tree Driven Theorem Solver (DT-Solver) by guiding the search procedure with state confidence and proof-level values. Specifically, DT-Solver introduces a dynamic-tree Monte-Carlo search algorithm, which dynamically allocates computing budgets for different state confidences, guided by a new proof-level value function to discover proof states that require substantial exploration. Experiments on two popular theorem-proving datasets, PISA and Mathlib, show significant performance gains by our DT-Solver over the state-of-the-art approaches, with a 6.65% improvement on average in terms of success rate. And especially under low computing resource settings (11.03% improvement on average).},
   booktitle={Proceedings of the 61st Annual Meeting of the Association for Computational Linguistics (Volume 1: Long Papers)},
	publisher={Association for Computational Linguistics},
	author={Wang, Haiming and Yuan, Ye and Liu, Zhengying and Shen, Jianhao and Yin, Yichun and Xiong, Jing and Xie, Enze and Shi, Han and Li, Yujun and Li, Lin and Yin, Jian and Li, Zhenguo and Liang, Xiaodan},
	year={2023}, month={Jul},
	pages={12632–12646}
}

@misc{bfs-prover,
      title={BFS-Prover: Scalable Best-First Tree Search for LLM-based Automatic Theorem Proving},
      author={Ran Xin and Chenguang Xi and Jie Yang and Feng Chen and Hang Wu and Xia Xiao and Yifan Sun and Shen Zheng and Kai Shen},
      year={2025},
      eprint={2502.03438},
      archivePrefix={arXiv},
      primaryClass={cs.AI},
      url={https://arxiv.org/abs/2502.03438},
}

@article{gelu,
  author       = {Dan Hendrycks and
                  Kevin Gimpel},
  title        = {Bridging Nonlinearities and Stochastic Regularizers with Gaussian
                  Error Linear Units},
  journal      = {CoRR},
  volume       = {abs/1606.08415},
  year         = {2016},
  url          = {http://arxiv.org/abs/1606.08415},
  eprinttype    = {arXiv},
  eprint       = {1606.08415},
  bibsource    = {dblp computer science bibliography, https://dblp.org}
}

@article{motivated-proofs,
   title={Motivated Proofs: what they are, why they matter and how to write them},
   volume={13},
   ISSN={1755-0211},
   url={http://dx.doi.org/10.1017/S1755020319000583},
   DOI={10.1017/s1755020319000583},
   number={1},
   journal={The Review of Symbolic Logic},
   publisher={Cambridge University Press (CUP)},
   author={MORRIS, REBECCA LEA},
   year={2019},
   month=nov, pages={23–46}
}

@article{graph4hol, title={Graph Representations for Higher-Order Logic and Theorem Proving}, volume={34}, url={https://ojs.aaai.org/index.php/AAAI/article/view/5689}, DOI={10.1609/aaai.v34i03.5689}, abstractNote={&lt;p&gt;This paper presents the first use of graph neural networks (GNNs) for higher-order proof search and demonstrates that GNNs can improve upon state-of-the-art results in this domain. Interactive, higher-order theorem provers allow for the formalization of most mathematical theories and have been shown to pose a significant challenge for deep learning. Higher-order logic is highly expressive and, even though it is well-structured with a clearly defined grammar and semantics, there still remains no well-established method to convert formulas into graph-based representations. In this paper, we consider several graphical representations of higher-order logic and evaluate them against the HOList benchmark for higher-order theorem proving.&lt;/p&gt;}, number={03}, journal={Proceedings of the AAAI Conference on Artificial Intelligence}, author={Paliwal, Aditya and Loos, Sarah and Rabe, Markus and Bansal, Kshitij and Szegedy, Christian}, year={2020}, month={Apr.}, pages={2967-2974} }

@misc{graph-premise-selection,
      title={GraphMind: Theorem Selection and Conclusion Generation Framework with Dynamic GNN for LLM Reasoning}, 
      author={Yutong Li and Yitian Zhou and Xudong Wang and GuoChen and Caiyan Qin},
      year={2025},
      eprint={2511.19078},
      archivePrefix={arXiv},
      primaryClass={cs.CL},
      url={https://arxiv.org/abs/2511.19078}, 
}

@misc{leanauto,
      title={Lean-auto: An Interface between Lean 4 and Automated Theorem Provers}, 
      author={Yicheng Qian and Joshua Clune and Clark Barrett and Jeremy Avigad},
      year={2025},
      eprint={2505.14929},
      archivePrefix={arXiv},
      primaryClass={cs.LO},
      url={https://arxiv.org/abs/2505.14929}, 
}

@misc{leansmt,
      title={Lean-SMT: An SMT tactic for discharging proof goals in Lean}, 
      author={Abdalrhman Mohamed and Tomaz Mascarenhas and Harun Khan and Haniel Barbosa and Andrew Reynolds and Yicheng Qian and Cesare Tinelli and Clark Barrett},
      year={2025},
      eprint={2505.15796},
      archivePrefix={arXiv},
      primaryClass={cs.LO},
      url={https://arxiv.org/abs/2505.15796}, 
}

@misc{deepseek-prover,
      title={DeepSeek-Prover: Advancing Theorem Proving in LLMs through Large-Scale Synthetic Data}, 
      author={Huajian Xin and Daya Guo and Zhihong Shao and Zhizhou Ren and Qihao Zhu and Bo Liu and Chong Ruan and Wenda Li and Xiaodan Liang},
      year={2024},
      eprint={2405.14333},
      archivePrefix={arXiv},
      primaryClass={cs.AI},
      url={https://arxiv.org/abs/2405.14333}, 
}

@article{leandojo,
  title={Leandojo: Theorem proving with retrieval-augmented language models},
  author={Yang, Kaiyu and Swope, Aidan and Gu, Alex and Chalamala, Rahul and Song, Peiyang and Yu, Shixing and Godil, Saad and Prenger, Ryan J and Anandkumar, Animashree},
  journal={Advances in Neural Information Processing Systems},
  volume={36},
  pages={21573--21612},
  year={2023}
}

@misc{minilang,
      title={A Minimalist Proof Language for Neural Theorem Proving over Isabelle/HOL}, 
      author={Qiyuan Xu and Renxi Wang and Peixin Wang and Haonan Li and Conrad Watt},
      year={2025},
      eprint={2507.18885},
      archivePrefix={arXiv},
      primaryClass={cs.PL},
      url={https://arxiv.org/abs/2507.18885}, 
}

@inproceedings{sledgehammer,
  title={Sledgehammer: judgement day},
  author={B{\"o}hme, Sascha and Nipkow, Tobias},
  booktitle={International Joint Conference on Automated Reasoning},
  pages={107--121},
  year={2010},
  organization={Springer}
}

@article{deBruijnNames,
title = {A Head-to-Head Comparison of de Bruijn Indices and Names},
journal = {Electronic Notes in Theoretical Computer Science},
volume = {174},
number = {5},
pages = {53-67},
year = {2007},
note = {Proceedings of the First International Workshop on Logical Frameworks and Meta-Languages: Theory and Practice (LFMTP 2006)},
issn = {1571-0661},
doi = {https://doi.org/10.1016/j.entcs.2007.01.018},
url = {https://www.sciencedirect.com/science/article/pii/S1571066107002319},
author = {Stefan Berghofer and Christian Urban}
}

@inproceedings{smtcoq,
  url       = "http://theory.stanford.edu/~barrett/pubs/EMT+17.pdf",
  author    = "Burak Ekici and Alain Mebsout and Cesare Tinelli and Chantal Keller and Guy Katz and Andrew Reynolds and Clark Barrett",
  title     = "{SMTC}oq: A Plug-In for Integrating {SMT} Solvers into {C}oq",
  booktitle = "Proceedings of the $29^{th}$ International Conference on Computer Aided Verification (CAV '17)",
  volume    = 10426,
  number    = 1,
  editor    = "Rupak Majumdar and Viktor Kuncak",
  pages     = "126--136",
  series    = "Lecture Notes in Computer Science",
  publisher = "Springer",
  month     = jul,
  year      = 2017,
  note      = "Heidelberg, Germany",
}

@article{coqhammer,
	author = {Czajka, {\L}ukasz and Kaliszyk, Cezary},
	date = {2018/06/01},
	doi = {10.1007/s10817-018-9458-4},
	id = {Czajka2018},
	isbn = {1573-0670},
	journal = {Journal of Automated Reasoning},
	number = {1},
	pages = {423--453},
	title = {Hammer for Coq: Automation for Dependent Type Theory},
	url = {https://doi.org/10.1007/s10817-018-9458-4},
	volume = {61},
	year = {2018}}

@misc{leanhammer,
      title={Premise Selection for a Lean Hammer}, 
      author={Thomas Zhu and Joshua Clune and Jeremy Avigad and Albert Qiaochu Jiang and Sean Welleck},
      year={2026},
      eprint={2506.07477},
      archivePrefix={arXiv},
      primaryClass={cs.LG},
      url={https://arxiv.org/abs/2506.07477}, 
}

\appendix
\subsection{Atomic Tactics}
\label{sec:Atomic Tactics}

Below is the list of all atomic tactics. $x \ll y$ means $y$ is type-dependent
on $x$. $\mathsf{unify}(x,y)$ holds if $x$ and $y$'s types can be unified, and
$\mathsf{transform}(x, f)$ transforms every subexpression of $x$ using the
operation described by $f$.  We first consider the tactics in
Figure~\ref{fig:Atomic Tactics from Lean}.  These are tactics already present in
Lean, with a few small modifications.  The value of $x$ supplied to the
$\mathsf{exact}$ tactic is only allowed to be a constant or free variable that
already appears in the goal (including its context), thus limiting the choice to
a finite set of possibilities. 

\begin{figure*}
	\begin{framed}
\[
	\mathsf{intro}
	\frac{\Goal{g} : \forall x:X.Y}
	{\Goal{g} := \uplambda x.\Goal{h}[x]; \{x : X\}\vdash\Goal{h} : Y} \qquad
	\mathsf{exact}(x)
	\frac{\Goal{g} : X, x : X, x\text{ is a constant or free}}
	{\Goal{g} := x}
\]
\[
	\mathsf{decide}
	\frac{\Goal{g} : X, X\text{ is decidable}}
	{\Goal{g} := \mathsf{decide}(X)} \qquad
	\mathsf{contradiction}
	\frac{\{x_i\} \vdash \Goal{g} : X, \lnot\Goal{g}\text{ contains of contradiction}}
	{\Goal{g} := \mathsf{contradict}(\Goal{g})}
\]
\[
	\mathsf{rfl}
	\frac{\Goal{g} : X = Y, \mathsf{unify}(X,Y)}
	{\Goal{g} := \mathsf{Eq.refl}} \qquad
	\mathsf{assumption}
	\frac{\{x_i:X_i\} \vdash \Goal{g} : X, \exists i.X_i = X}
	{\Goal{g} := x_i}
\]
\[
	\mathsf{apply}(f)
	\frac{\Goal{g}:Y, f:X_1 \to \cdots \to X_n; \exists i.\mathsf{unify}(X_{i+1} \to \cdots \to X_n,Y)}
	{\Goal{g} := f\,\Goal{g}_1,\dots,\Goal{g}_i;
	\Goal{g}_j:X_j}
\]
\[
	\mathsf{cases}(m)
	\frac{\{m: M\} \vdash \Goal{g} : X}
	{
	\Goal{g} := X\mathsf{.rec}(\func x_c.\Goal{g}_c(x_C) \mid c);
	\{x_c \vdash \Goal{g}_c : X | c\text{ is a constructor of }M\}}
\]
\[
	\mathsf{induction}(m,r)
	\frac{\{m: M\} \vdash \Goal{g} : X_m}
	{
	\Goal{g} := r(\func x \func X_{c(x)}.\Goal{g}_c(x_c) \mid c);
	\{x_c,h:X_{c(x)} \vdash \Goal{g}_c : X | c\text{ is a minor of }r\}}
\]
\[
	\mathsf{revert}(i)
	\frac{\{x_1:X_1,\dots,x_n:X_n\}\Goal{g} : Y, x_i \ll x_j\text{ for }j > i}
	{\Goal{g} := \Goal{h}\,x_i,\dots,x_n;\{x_1,\dots,x_{i-1}\} \vdash \Goal{h} :
	X_i \to \cdots \to X_n \to Y}
\]
	\end{framed}
	\caption{Built-in Lean tactics as atomic tactics}
	\label{fig:Atomic Tactics from Lean}
\end{figure*}

We create the following synthetic atomic tactics to handle edge cases:

\begin{figure*}
	\begin{framed}
\[
	\mathsf{pi}\frac{\Goal{y} : \mathsf{Type}}{
		\Goal{y} := \forall x:\Goal{X}.\Goal{z},
		\Goal{X} : \mathsf{Type},
		\Goal{z}[x : \Goal{X}] : \mathsf{Type}
	} \qquad
	\mathsf{inhabit}\frac{\Goal{g} : X, g \in \mathsf{Inhabited}}{
		\Goal{g} := \mathsf{default}(X)
	}
\]
\[
	\mathsf{reduceBeta}
	\frac{\Goal{g} \text{ has a \textbeta-reducible subexpression}}
	{\mathsf{transform}(\Goal{g}, \beta\text{-reduce})}
\]
\[
	\mathsf{normalize}
	\frac{\Goal{g} \text{ has type class calls }}
	{\mathsf{transform}(\Goal{g}, \text{unfold type class})}
\]
\[
	\mathsf{tailArg}\frac{\Goal{y} : Y}{
		\Goal{y} := \Goal{f}\,\Goal{x},
		\Goal{X} : \mathsf{Type},
		\Goal{f} : \Goal{X} \to Y,
		\Goal{x} : \Goal{X}
	}
\]
\[
	\mathsf{motivatedApply}(f)
		\frac{\Goal{g} : Y, f: (X_1,\dots,X_n) \to B}{
			\Goal{g} := \mathsf{Eq.mp}\,\Goal{c}\,(f\,\Goal{x}_1,\dots,\Goal{x_n}),
			\Goal{x_i} : X_i,
			\Goal{c} : B = Y
		}
\]
\[
	\mathsf{congruenceArg}
	\frac{\Goal{g}: b_1 = b_2, b_1,b_2 : \beta}
	{\Goal{g}:= \mathsf{congrArg} \Goal{\alpha}\,\Goal{a_1}\,\Goal{a_2}\,\Goal{f}\,\Goal{h};
	\Goal{a_1},\Goal{a_2}: \Goal{\alpha},
	\Goal{f}: \Goal{\alpha} \to \beta,
	\Goal{h}: \Goal{a_1} = \Goal{a_2}}
\]
\[
	\mathsf{congruenceFun}
	\frac{\Goal{g}: b_1 = b_2, b_1,b_2 : \beta}
	{\Goal{g}:= \mathsf{congrFun} \Goal{\alpha}\,\Goal{a}\,\Goal{f_1}\,\Goal{f_2}\,\Goal{h};
	\Goal{a}: \Goal{\alpha},
	\Goal{f_1},\Goal{f_2}: \Goal{\alpha} \to \beta,
	\Goal{h}: \Goal{f_1} = \Goal{f_2}}
\]
	\end{framed}
	\caption{Synthetic atomic tactics for handling edge cases}
	\label{fig:Atomic Tactics Synthetic}
\end{figure*}

Figure~\ref{fig:Atomic Tactics Synthetic} describes a set of hand-crafted
tactics which, together with those in Figure~\ref{fig:Atomic Tactics from Lean}
and the two positional tactics below, complete our set of atomic
tactics.  These are crafted to help meet our objectives of completeness and
limited cross-section. Each rule has a finite number of parameters, each of which
has a finite number of choices. This provides a finite action space for any
agent using these atomic tactics.
$\mathsf{pi}$ generates a raw $\forall$ binder; $\mathsf{inhabit}$ generates a
default value for a target which is a known inhabited type;
$\mathsf{reduceBeta}$ \textbeta-reduces all subexpressions in the current goal;
$\mathsf{normalize}$ unfolds type class function calls in Lean, which is a
mechanism for handling arithmetic operators; $\mathsf{motivatedApply}$ is an
apply-like tactic which enforces type equality not via aggressive unification
but rather via a \emph{conduit}
$\Goal{c}$;
This is used as a last-resort mechanism when converting a proof to atomic tactics, if
the $\mathsf{apply}$ tactic is unusable because of its reliance on unification.
$\mathsf{tailArg}$ introduces a raw unary function application;
$\mathsf{congruenceArg}$ and $\mathsf{congruenceFun}$ are last-resort
tactics for handling congruences that cannot be written as rewrites.


There are two last rules in our atomic set, which are more easily explained
using examples.
\lstinline{rewritePos(heq, locus, symm)} is a rewrite tactic using a lemma
	$h$ which acts on a subexpression position called a \keyword{locus} rather than
	the $n$th match of $h$. For example,
	\[
		\mathsf{rewritePos}(h, \Pos{p}, \leftarrow)
		\frac{\vdash g(\mask{b + 1}) = f(a), h: a = b + 1}
		{\vdash g(a) = f(a)}
	        \]
Here, $\Pos{p}$ points to the expression $b+1$, and $\leftarrow$ indicates that
the equation $a=b+1$ is used as a rewrite rule from right to left.

A similar rule using positions is \lstinline{generalizeAt(locus)}, which
replaces some subexpression with a fresh variable.
	For example,
	\[
		\mathsf{generalizeAt}(\Pos{p})
		\frac{\vdash f'(\mask{2x + 3}) = 0}
		{\{y : \mathbb R\} \vdash f'(y) = 0}
	\]
Here, we replace the subexpression $2x+3$ with a fresh variable $y$.

The positional tactics $\mathsf{rewritePos}$ and $\mathsf{generalizeAt}$ act
directly on a specific subtree. This enables neural networks to
focus on a subexpression using attention-like mechanisms and skip over the
difficult problem of expression navigation using $\mathsf{conv}$ or
$\mathsf{congr}$.

\subsection{Example of Transposing Atomization}

\begin{figure}
\centering
\begin{subfigure}{.5\textwidth}
	\centering
	\includegraphics[width=.9\textwidth]{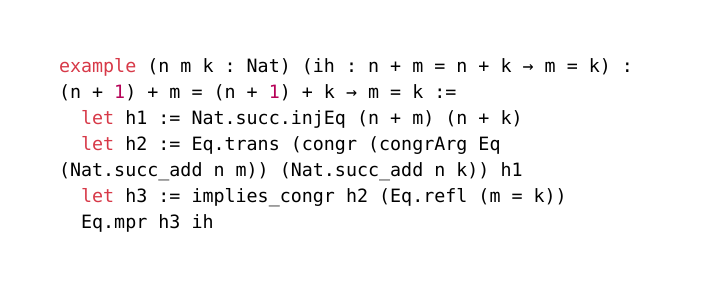}
	\caption{Before atomization}
\end{subfigure}

\begin{subfigure}{.5\textwidth}
	\centering
	\includegraphics[width=.9\textwidth]{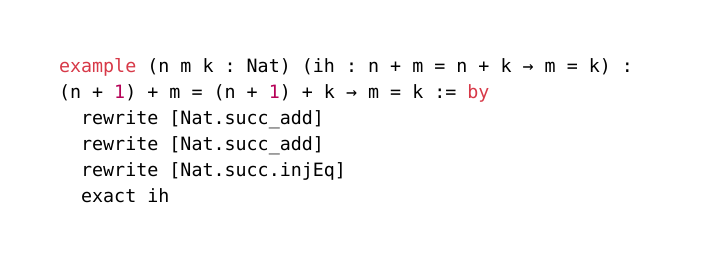}
	\caption{After atomization}
\end{subfigure}
\caption{Example of a proof before and after atomization}
	\label{fig:Example of atomization}
\end{figure}

In Figure~\ref{fig:Example of atomization}, we show an example of an atomized
proof. The atomized proof contains rewrite tactics that have exactly one operand
and applies to a single locus.

\subsection{Evaluation Result on Stdlib}
\label{sec:Evaluation Result on Stdlib}

Evaluation result of Nazrin on Lean's standard library is in
Figure~\ref{fig:Evaluation result on stdlib}. We train Nazrin on successive
slices of Lean Standard Library and Mathlib 4, using topological ordering to
ensure the absence of information leakage.

\begin{figure}
		\centering
		\includesvg[width=.9\linewidth]{diagrams/result/train-mathlib.svg}
		\caption{Training history of Nazrin prover on slices of Mathlib. After
	training on slice 1, we evaluate it on slice 2}
\end{figure}
\begin{figure}
		\centering
		\includesvg[width=.9\linewidth]{diagrams/result/search-stdlib-s2.svg}
		\caption{Search evaluation of the model on epoch 100 on slice 2 with a
	time limit of 15 seconds per theorem;  Blue dots indicate successes, and orange
	dots indicate failures.}
	\label{fig:Evaluation result on stdlib}
\end{figure}

\end{document}